\documentclass[sigconf, nonacm]{acmart}
\usepackage{amsmath}
\usepackage{amsfonts}
\usepackage{graphicx}
\usepackage{subcaption}
\usepackage{comment}
\usepackage{siunitx}
\begin{document}

\title{Perplexity-Homophily Index: Homophily through Diversity in Hypergraphs}


\author{Gaurav Kumar}
\orcid{0009-0007-9103-794X}
\affiliation{%
  \institution{IISER Pune} 
  \country{India}
}
\email{gaurav2k29@gmail.com}

\author{Akrati Saxena}
\affiliation{%
  \institution{LIACS, Leiden University}
  \country{The Netherlands}}
  \email{a.saxena@liacs.leidenuniv.nl}

\author{Chandrakala Meena}
\affiliation{%
  \institution{IISER Pune}
  \country{India}}
  \email{chandrakala@iiserpune.ac.in}






\renewcommand{\shortauthors}{Kumar et al.}

\begin{abstract}

Real-world complex systems are often better modeled as hypergraphs, where edges represent group interactions involving multiple entities. Understanding and quantifying homophily (similarity-driven association) in such networks is essential for analyzing community formation and information flow. We propose a hyperedge-centric framework to quantify homophily in hypergraphs. Each interaction is represented as a hyperedge, and its interaction perplexity measures the effective number of distinct attributes it contains. Comparing this observed perplexity with a degree-preserving random baseline defines the diversity gap, which quantifies how diverse an interaction is than expected by chance. The global homophily score for a network, called ``Perplexity-Homophily Index", is computed by averaging the normalized diversity gap across all hyperedges. 
Experiments on synthetic and real-world datasets show that the proposed index captures the full distribution of homophily and reveals how homophilic and heterophilic tendencies vary with interaction size in hypergraphs.
\end{abstract}



\maketitle

\section{Introduction}
Traditionally, complex systems are modeled as graphs with edges capturing pairwise interactions. However, many real-world interactions involve multiple nodes (groups) rather than pairs, such as co-authors on a paper, items bought together, or legislators co-sponsoring a bill. These systems are better captured by hypergraphs, where hyperedges represent group interactions. Real-world networks, driven by human behavior, are evolved using homophily \cite{mcphersonBirdsFeatherHomophily2001} that shows the tendency of entities to interact preferentially with others sharing similar attributes or interests (e.g., age, beliefs, nationality). Measuring homophily is essential for understanding community formation, information flow \cite{golubHowHomophilyAffects2009, rai2025ipsr}, and the emergence of structural inequalities in social systems \cite{saxena2024fairsna}. 


A widely used measure of homophily in pairwise graphs is Newman's assortativity coefficient \cite{newmanMixingPatternsNetworks2003}, which quantifies the deviation of a graph's structure from random mixing and enables comparisons across networks. Despite extensive work on homophily in pairwise graphs \cite{saxena2025homophily}, such measures for hypergraphs remain limited. 
Veldt et al. \cite{veldtCombinatorialCharacterizationsImpossibilities2023} characterize the homophily types for a specific class (e.g., majority or minority preference) but provide no aggregate score for the whole hypergraph and assumes a $k$-uniform hypergraph. Li et al. \cite{liWhenHypergraphMeets2025} estimate node and hyperedge homophily by counting the fraction of same-class pairs and aggregating it for the whole hypergraph, while Telyatnikov et al. \cite{telyatnikovHypergraphNeuralNetworks2023} use message passing to define node-level homophily. However, all existing approaches lack an appropriate null model, and their scores are not directly comparable across hypergraphs. 

In this work, we introduce a novel, hyperedge-centric framework to measure homophily in hypergraphs. We first define the interaction perplexity \(D(e)\), which quantifies the ``effective number of distinct attributes'' within a hyperedge. Perplexity is an order-\(1\) Hill number \cite{hillDiversityEvennessUnifying1973} originally used to measure the diversity of species in a region. 
In hypergraphs, hyperedges can vary from pure (all nodes have the same attribute) to balanced (attributes are evenly distributed). Interaction perplexity captures this full spectrum of hyperedge compositions. We compare the observed perplexity of each hyperedge to a random baseline that accounts for both the global attribute distribution and the degree distribution. The homophily score \(\phi(e)\) for a hyperedge \(e\) is defined as the normalized difference of the observed and the baseline perplexity. 
The Perplexity-Homophily Index for a hypergraph, \(\Phi(H)\), is the average over homophily scores of all hyperedges. This index is comparable across datasets with differing hyperedge sizes and attribute distributions. 

We validate our method on various real-world hypergraphs, and the results show that $\Phi(H)$ is able to characterize homophily across a wide range of datasets. Moreover, we highlight how homophilic or heterophilic tendencies vary with interaction size, such as the stronger homophily in small co-sponsorship groups and near-random mixing in larger ones for US Congressional bill co-sponsorship dataset.

\section{The Proposed Homophily Measure}\label{sec:diversity}
A hypergraph $H$ is represented as $(V, E)$, where $V$ and $E$ are the set of nodes and hyperedges, respectively. A hyperedge $e \in E$ is a non-empty subset of $V$ and every node $v \in V$ is associated with an attribute from a set of possible attributes $M$. 
Based on the distribution of attributes of nodes in a hyperedge, it can classified as pure if all nodes have the same attribute, balanced if all attributes are in equal proportion, and biased if one or more attributes dominate. 
A $k$-uniform hypergraph is one where each hyperedge has exactly $k$ nodes.

\subsection{Interaction Perplexity}\label{sec:diversity:perplexity}
Let $p_{i,e}$ denote the fraction of nodes with attribute $i$ in hyperedge $e$. We now define the interaction perplexity $D(e)$ as the perplexity calculated over the distribution of attributes $\{p_{i,e}\}_{i\in M}$:
\begin{equation} 
    D(e) = 2^{- \sum_{i \in M} p_{i,e} \log_2 p_{i,e}}
    \label{eq:perplexity}
\end{equation}

$D(e)$ ranges from $1$ to the number of attributes in $e$, $m_e$. A pure interaction will have $D(e)=1$, while a balanced interaction will have $D(e)=m_e$. Biased interactions will have a perplexity between $1$ and $m_e$. Thus, $D(e)$ captures the full spectrum of attribute dominance within $H$, which is essential for accurately quantifying the homophily of the hypergraph.

\textbf{Example.} Consider a hyperedge with four researchers from three departments: Physics $(2)$, CS $(1)$, and Math $(1)$, giving the distribution $(2,1,1)$. While three departments are present, Physics dominates, resulting in lower effective diversity. The interaction perplexity for this distribution is $2.83$, meaning the group behaves as if it effectively contains about $2.83$ equally represented departments. If all researchers were from Physics, the perplexity would be $1$, and if each belonged to a different department, it would be $4$. 

\begin{figure}[H]
    \centering
    \includegraphics[width=0.5\linewidth]{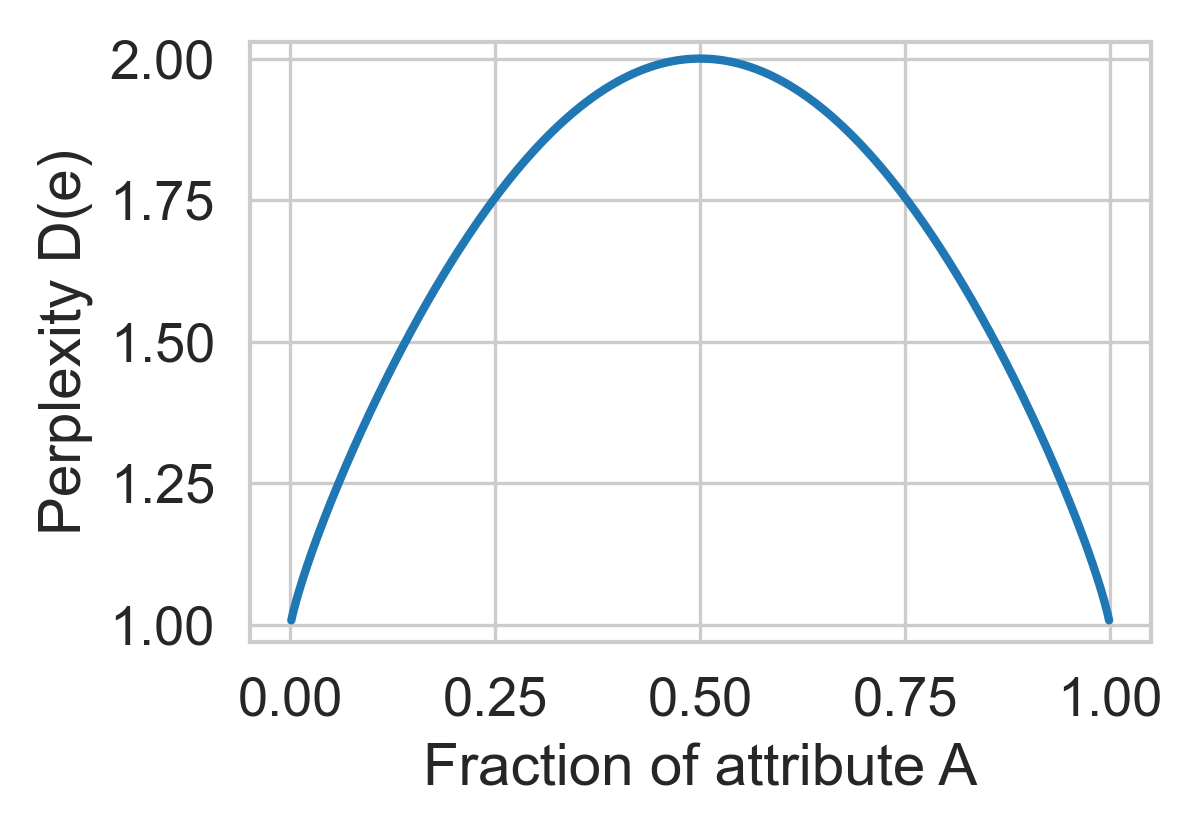}
    \caption{
        Perplexity $D(e)$ for a hyperedge with two attributes $A$ and $B$.
    }
    \label{fig:perplexity-two-attributes}
\end{figure}

\subsection{Baseline and Diversity Gap}\label{sec:diversity:baseline}
To assess homophily, the observed perplexity must be compared to a null model that accounts for both the global attribute distribution and degree distribution of nodes. In a degree-preserving randomized $k$-uniform hypergraph, the baseline perplexity $B_k$ represents the expected number of distinct attributes in a hyperedge of size $k$ under random mixing.

A hypergraph with a highly skewed global attribute distribution (e.g., 90\% of nodes sharing one attribute) will naturally have a low average interaction perplexity, since most interactions will be dominated by the majority group. Similarly, high-degree nodes are more likely to appear in sampled interactions, increasing the expected prevalence of their attributes. 

We estimate the baseline using a Monte Carlo procedure: for each sample, we draw $k$ nodes without replacement, with sampling probability proportional to their $k$-degree (the number of size-k hyperedges the node belongs to), and then compute the interaction perplexity of the resulting hyperedge. For a nonuniform hypergraph, each $k$-uniform sub-hypergraph will have its own baseline. 
The underlying distribution of the null model is a multivariate noncentral hypergeometric distribution. Computing the expected value of the baseline perplexity for a hypergraph would require summing over a large state space of all possible combinations of attributes, which is not feasible as the size of the hypergraph increases. 

We define the Diversity Gap $\Delta(e)$ of an interaction as the difference between the observed perplexity and its baseline.
\begin{equation*} 
    \Delta(e) = B_{|e|} - D(e)
\end{equation*}

When $\Delta(e)>0$, the observed hyperedge is less diverse than the expected diversity (high homophily). Similarly, $\Delta(e)<0$ shows that the hyperedge is more diverse than expected (heterophily). A gap near zero suggests that the diversity of the hyperedge is as expected, and interactions are consistent with the null model.


The Diversity Gap is maximum for a pure hyperedge, since $D(e)=1$:
\begin{equation*}
    \Delta_\text{max}(e) = B_{|e|} - 1
\end{equation*}
and minimum for a balanced hyperedge:
\begin{equation*}
    \Delta_\text{min}(e)= B_{|e|}-m_e
\end{equation*}

\subsection{Perplexity-Homophily Index}\label{sec:diversity:homophily}
To obtain a homophily score that is comparable across hyperedges, we normalize the diversity gap by its maximum value: 
\begin{equation} 
    \phi(e) = \frac{\Delta(e)}{\Delta_\text{max}(e)} = \frac{B_{|e|} - D(e)}{B_{|e|} - 1}
\end{equation}
This normalization provides a clear interpretation for measuring homophily, as it represents the fraction of the maximum possible reduction in diversity relative to the baseline. The maximum score of $+1$ corresponds to a fully pure hyperedge, while $\phi(e)=0.8$ means that the hyperedge is $80\%$ less diverse than expected, or $80\%$ of the way being completely homophilic. The minimum score depends on the number of attributes in the hyperedge, $\phi_\text{min}(e)=\Delta_\text{min}(e)/\Delta_\text{max}(e)$, providing a meaningful way to compare homophily across hyperedges with different attribute compositions.

The Perplexity-Homophily Index of a hypergraph $H$ is defined as the average homophily score over all hyperedges: 
\begin{equation}
    \Phi(H) = \frac{1}{|E|}\sum_{e\in E}\phi(e)
\end{equation}

Similarly, $\Phi(H_k)$ measures the homophily of $k$-uniform sub-hypergraphs ($H_k$) present in a hypergraph $H$ (Fig.~\ref{fig:phi-vs-k}). For $k=2$, the index closely approximates Newman's assortativity coefficient.

\section{Results}

\begin{figure}[b]
    \centering
    \begin{subfigure}[b]{0.48\columnwidth}
        \centering
        \includegraphics[width=\textwidth]{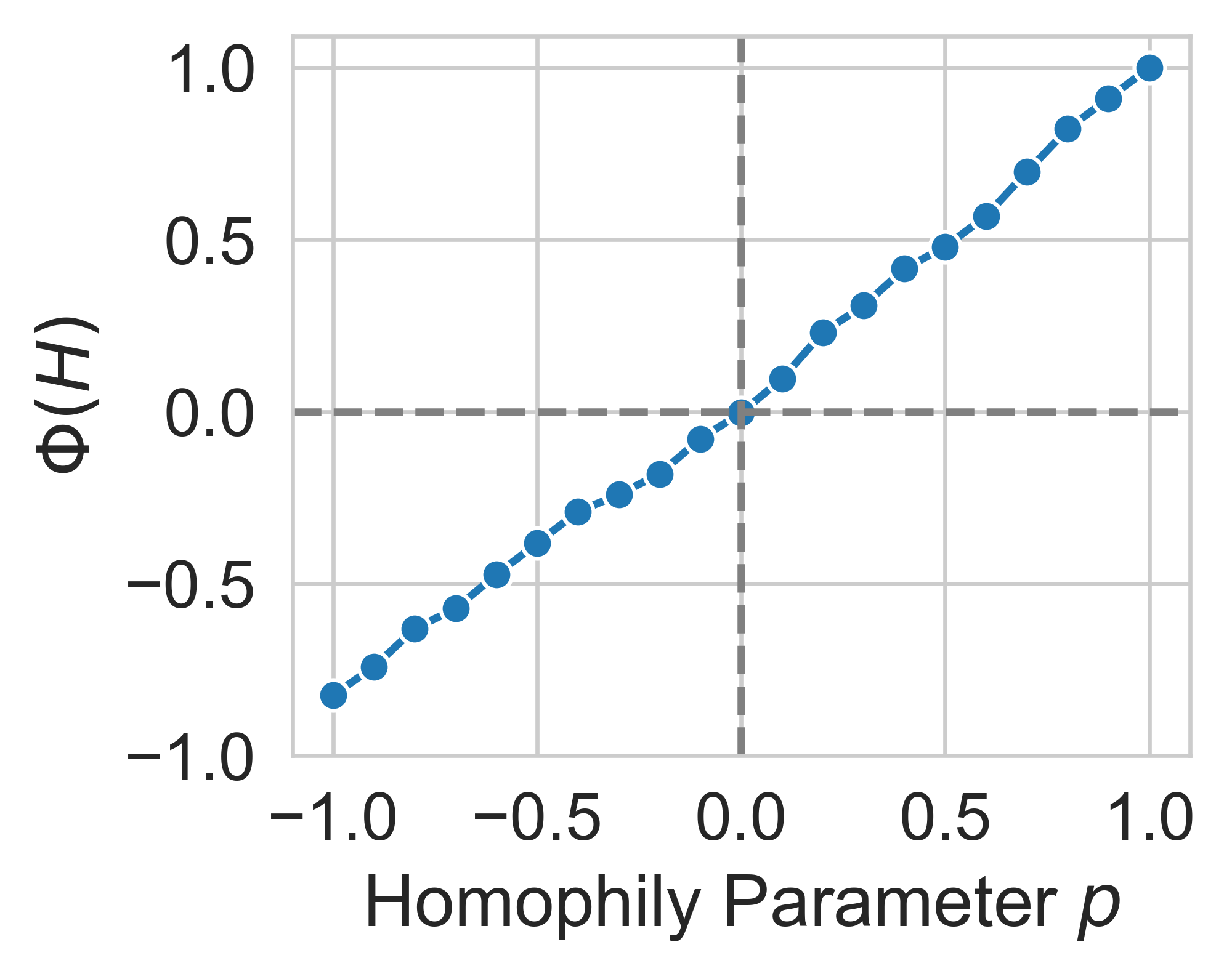}
        \caption{$\Phi(H)$ vs $p$}
        \label{fig:synthetic:h_vs_p}
    \end{subfigure}
    \hfill
    \begin{subfigure}[b]{0.48\columnwidth}
        \centering
        \includegraphics[width=\textwidth]{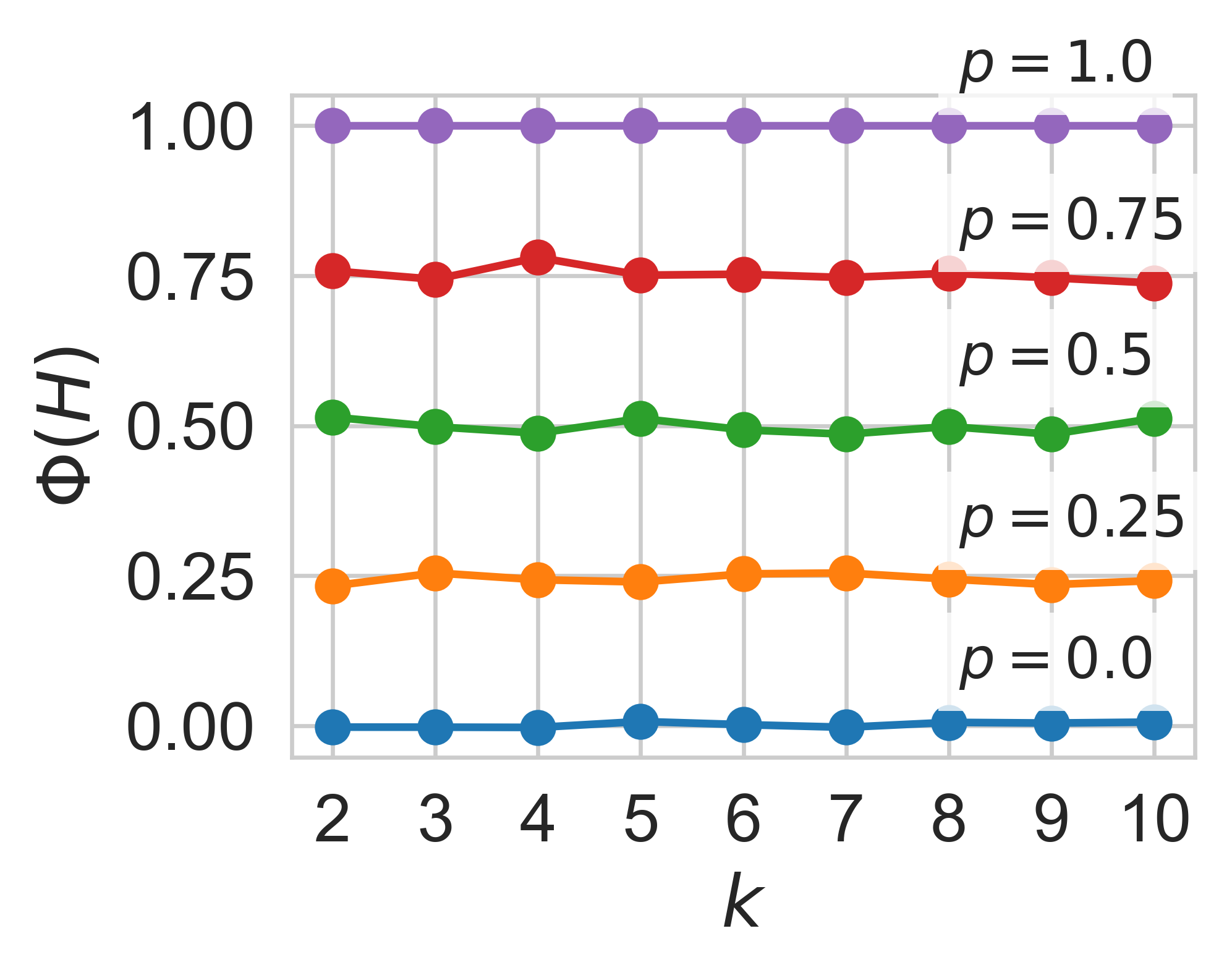}
        \caption{$\Phi(H)$ vs $k$ for multiple $p$}
        \label{fig:synthetic:h_vs_k}
    \end{subfigure}

    \caption{(a). Homophily vs. $p$ for $10$-uniform hypergraph with $1000$ nodes and $10$ evenly distributed attributes, and (b). Comparison of homophily scores for $k$-uniform hypergraphs with varying $p$ on $1000$ nodes network with $10$ evenly distributed attributes.}
    \label{fig:synthetic}
\end{figure}

\begin{table*}[t]
    \caption{Summary of real-world hypergraph datasets used in the analysis.}
    \vspace{-.5em}
    \label{tab:datasets}
    \centering
    \begin{tabular}{l l r l r l r r}
    \toprule
     \textbf{Dataset ($H$)} & \textbf{Nodes} & $|V|$ & \textbf{Hyperedges} & $|E|$ & \textbf{Attribute} & $|M|$ & $\Phi(H)$ \\
    \midrule
    {Walmart Trips} \cite{Amburg-2020-categorical} & Products & \num{88860} & Shopping carts & \num{65979} & Product category & \num{11} & \num{0.47} \\
    {Trivago Clicks} \cite{chodrow2021hypergraph} & Hotels & \num{172738} & Browsing sessions & \num{220971} & Country & \num{160} & \num{0.98} \\
    {Primary School} \cite{chodrow2021hypergraph} & Students/Teachers & \num{242} & Proximity groups & \num{12704} & Class / Teacher & \num{11} & \num{0.43} \\ 
    {High School} \cite{chodrow2021hypergraph, Mastrandrea-2015-contact} & Students & \num{327} & Proximity groups & \num{7818} & Class & \num{9} & \num{0.73} \\
    {US House Bills} \cite{chodrow2021hypergraph, Fowler-2006-connecting} & Legislators & \num{1494} & Bills & \num{54933} & Political party & \num{2} & \num{0.32} \\
    {US Senate Bills} \cite{chodrow2021hypergraph, Fowler-2006-connecting} & Legislators & \num{294} & Bills & \num{21721} & Political party & \num{2} & \num{0.24} \\ 
    {US House Committees} \cite{chodrow2021hypergraph, house-stewart} & Legislators & \num{1290} & Committees & \num{336} & Political party & \num{2} & \num{-0.03} \\
    {US Senate Committees} \cite{chodrow2021hypergraph, house-stewart} & Legislators & \num{282} & Committees & \num{301} & Political party & \num{2} & \num{-0.05} \\
    \bottomrule
    \end{tabular}
\end{table*}

\begin{figure*}[!t]
    \centering
    \begin{subfigure}[t]{0.24\textwidth}
        \centering
        \includegraphics[width=\linewidth]{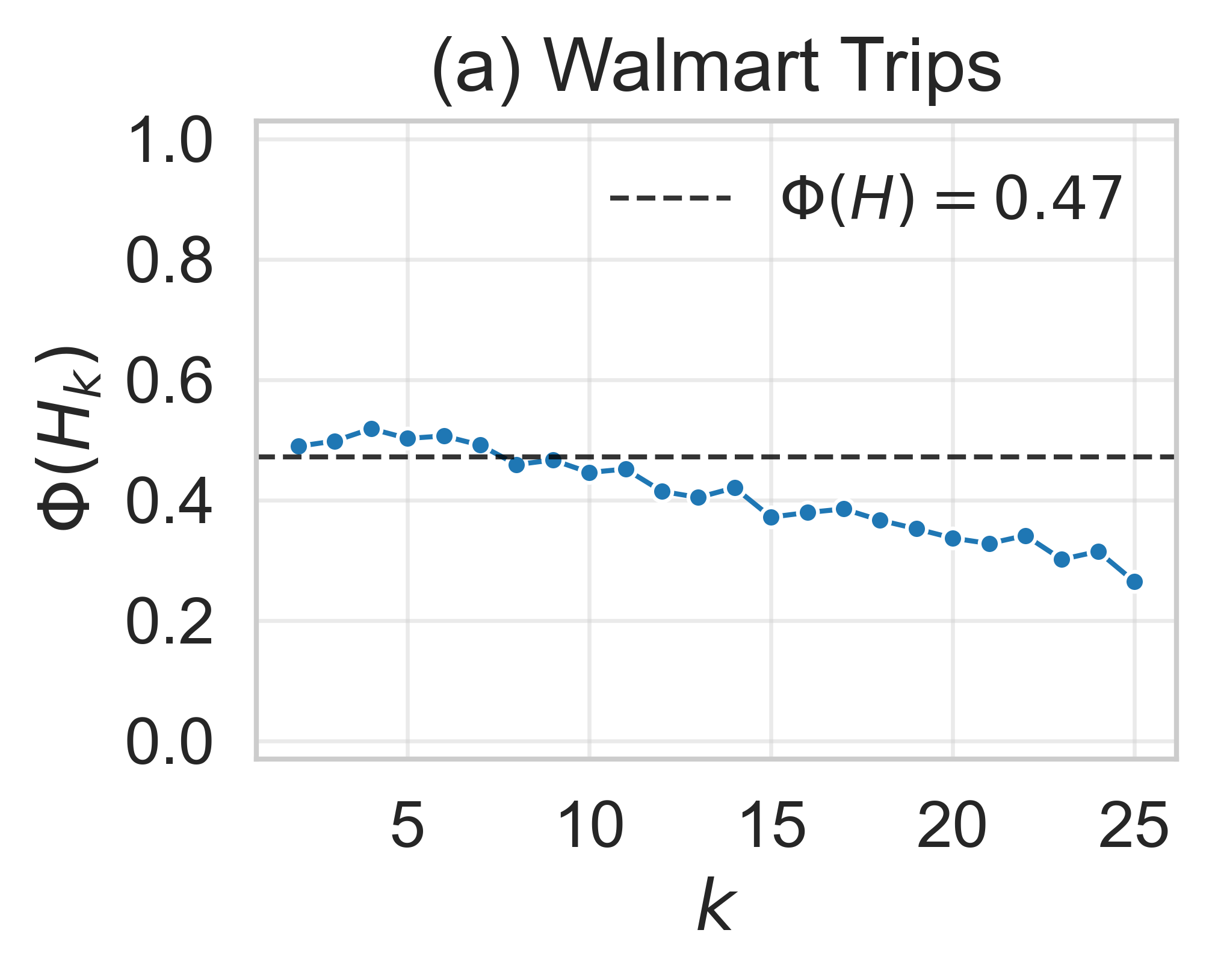}
    \end{subfigure}
    \begin{subfigure}[t]{0.24\textwidth}
        \centering
        \includegraphics[width=\linewidth]{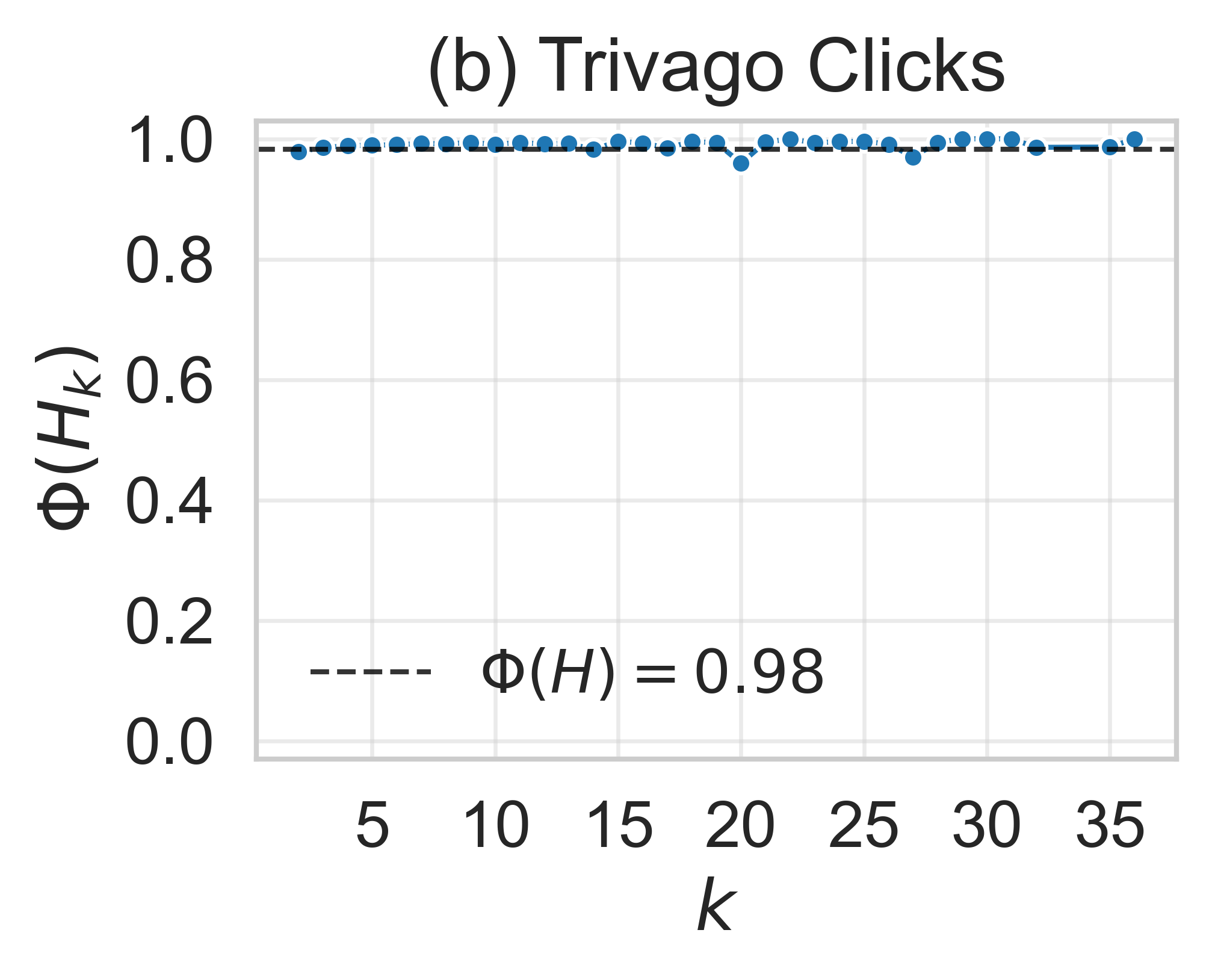}
    \end{subfigure}
    \begin{subfigure}[t]{0.24\textwidth}
        \centering
        \includegraphics[width=\linewidth]{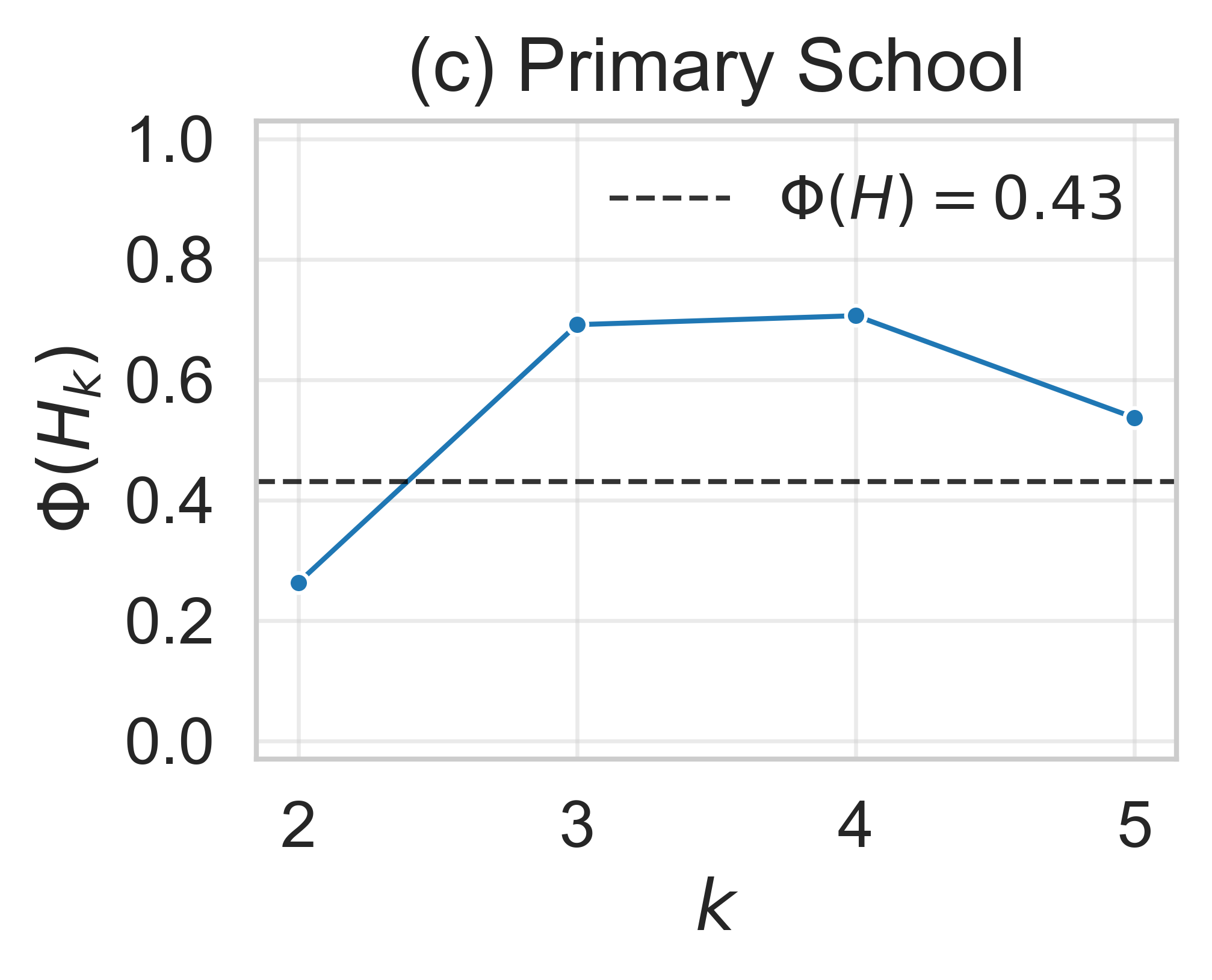}
    \end{subfigure}
    \begin{subfigure}[t]{0.24\textwidth}
        \centering
        \includegraphics[width=\linewidth]{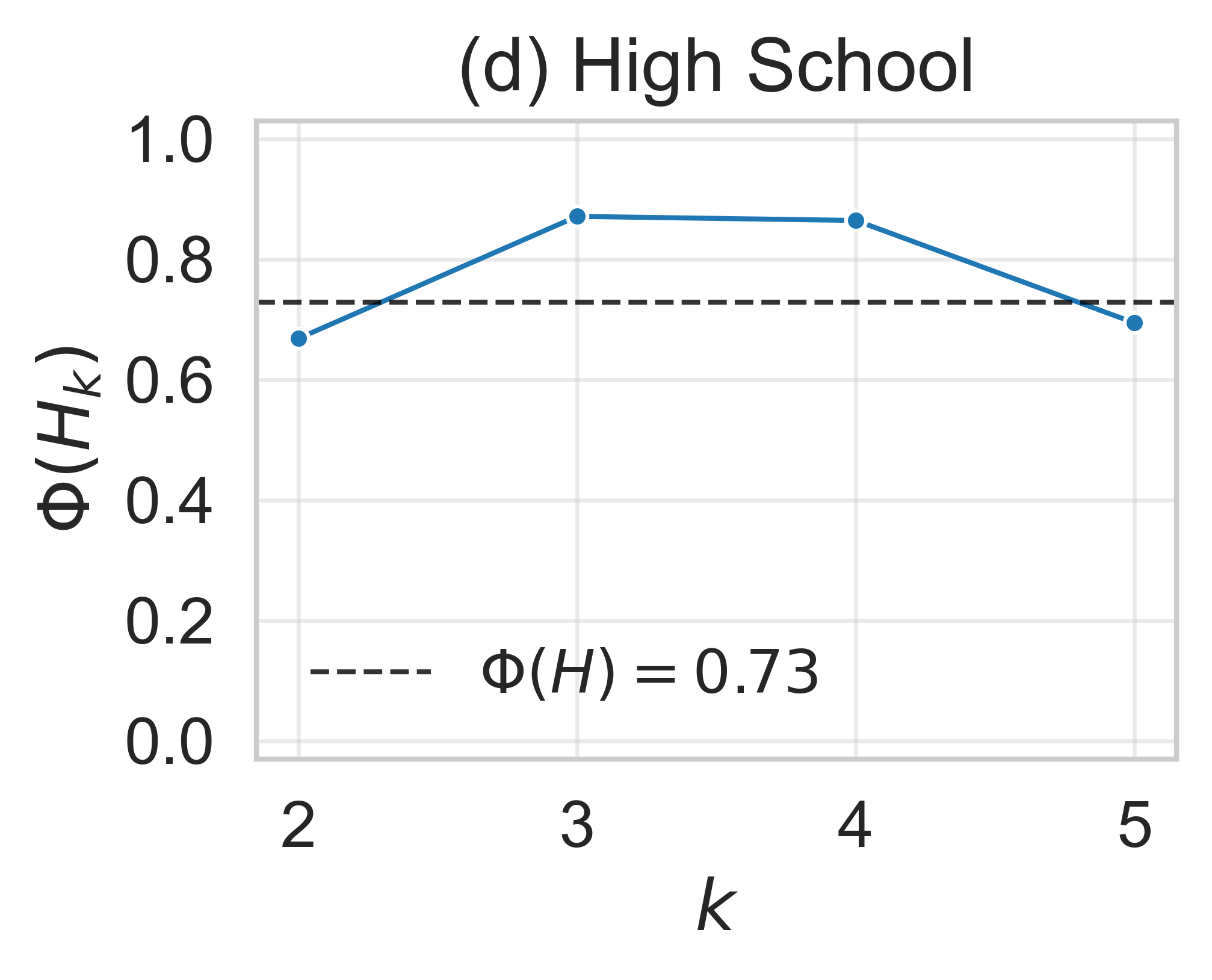}
    \end{subfigure}
    \begin{subfigure}[t]{0.24\textwidth}
        \centering
        \includegraphics[width=\linewidth]{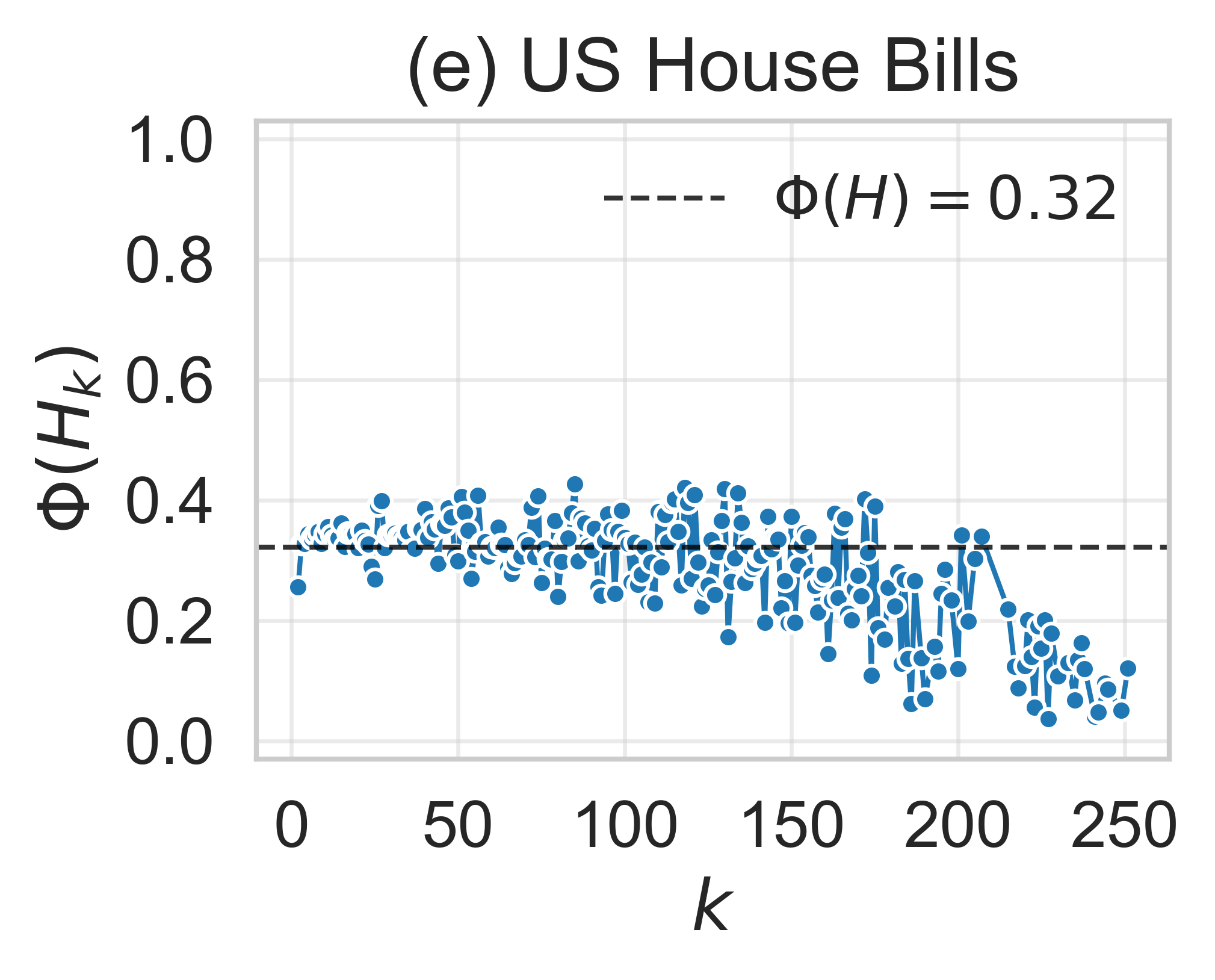}
    \end{subfigure}
    \begin{subfigure}[t]{0.24\textwidth}
        \centering
        \includegraphics[width=\linewidth]{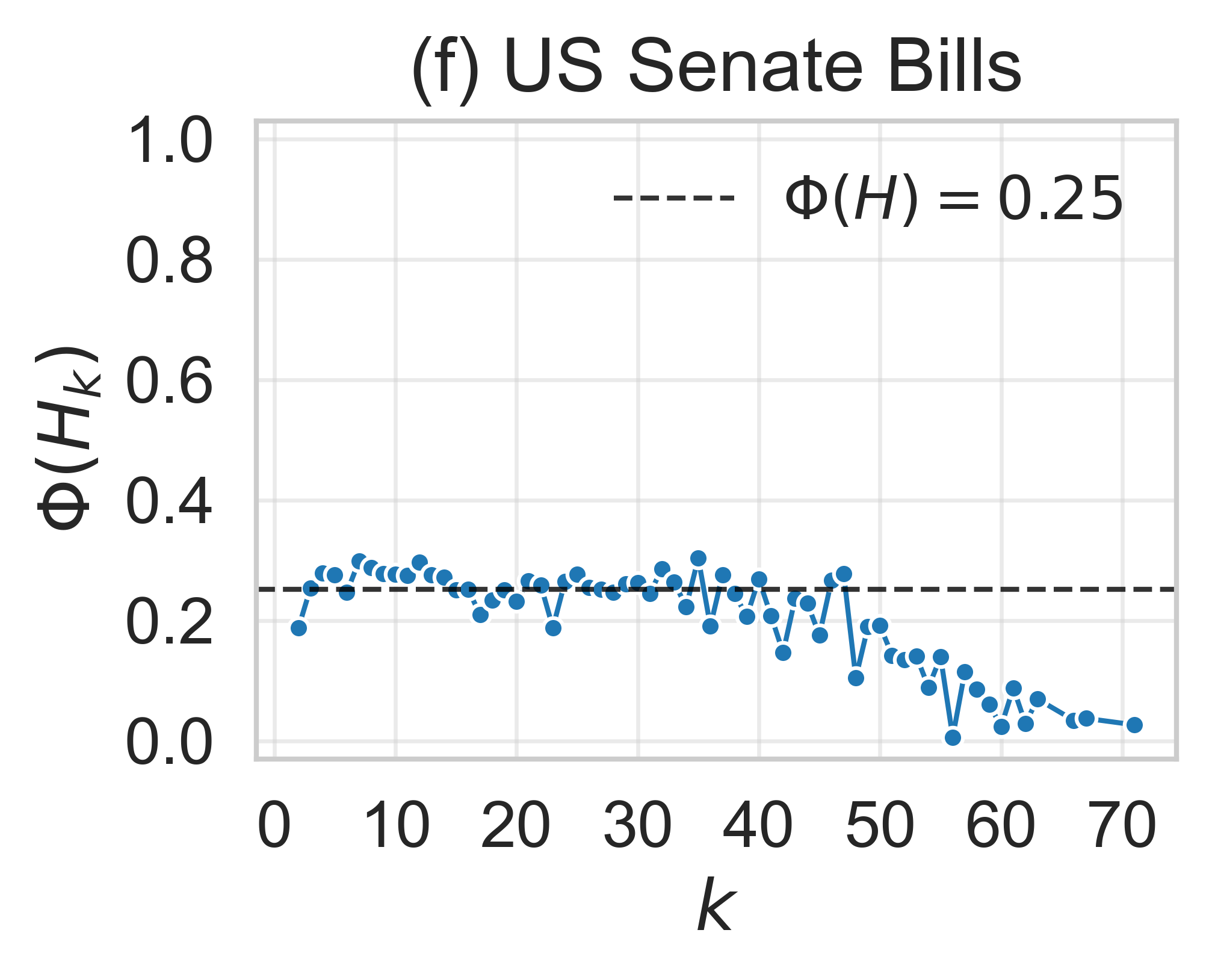}
    \end{subfigure}
    \begin{subfigure}[t]{0.24\textwidth}
        \centering
        \includegraphics[width=\linewidth]{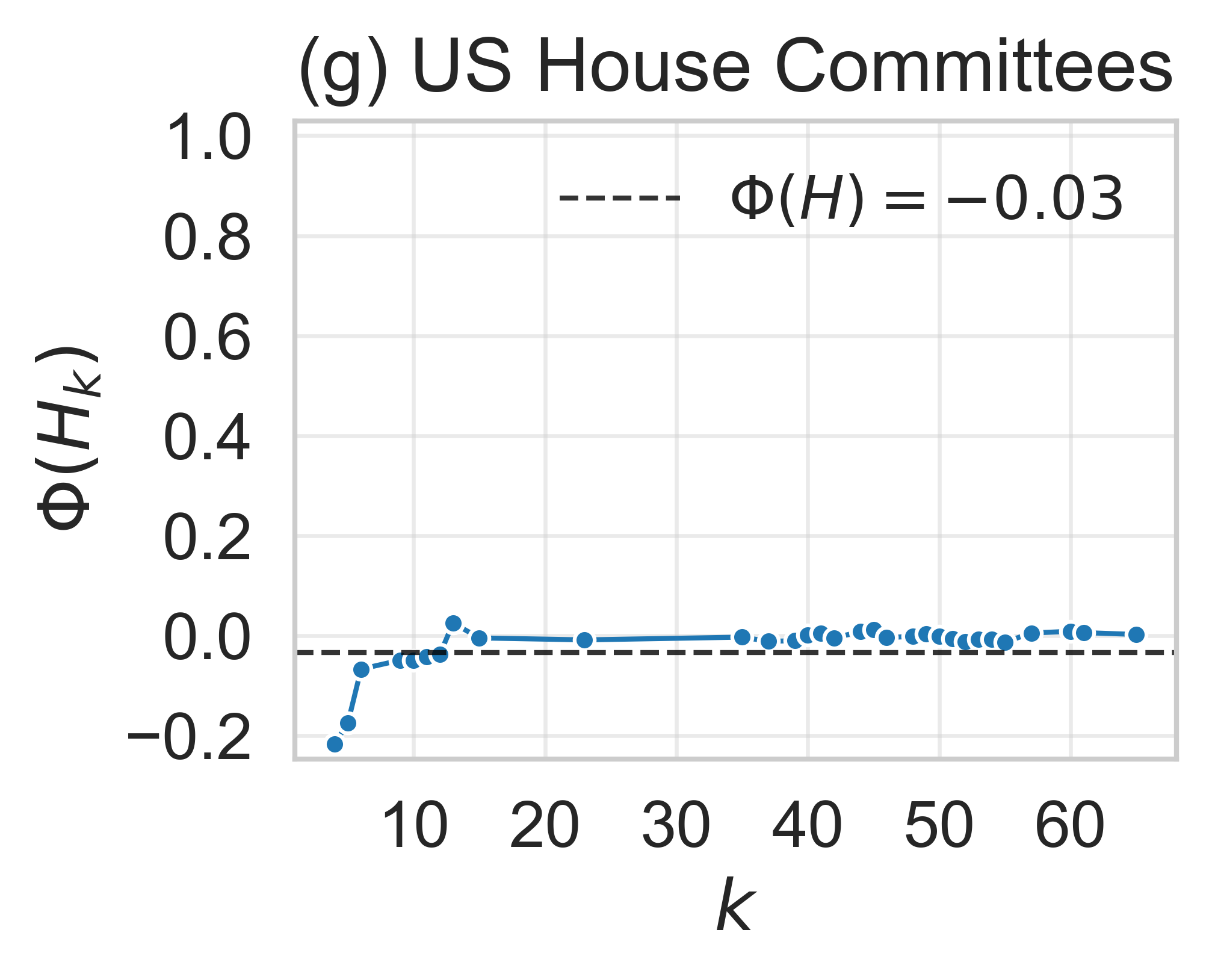}
    \end{subfigure}
    \begin{subfigure}[t]{0.24\textwidth}
        \centering
        \includegraphics[width=\linewidth]{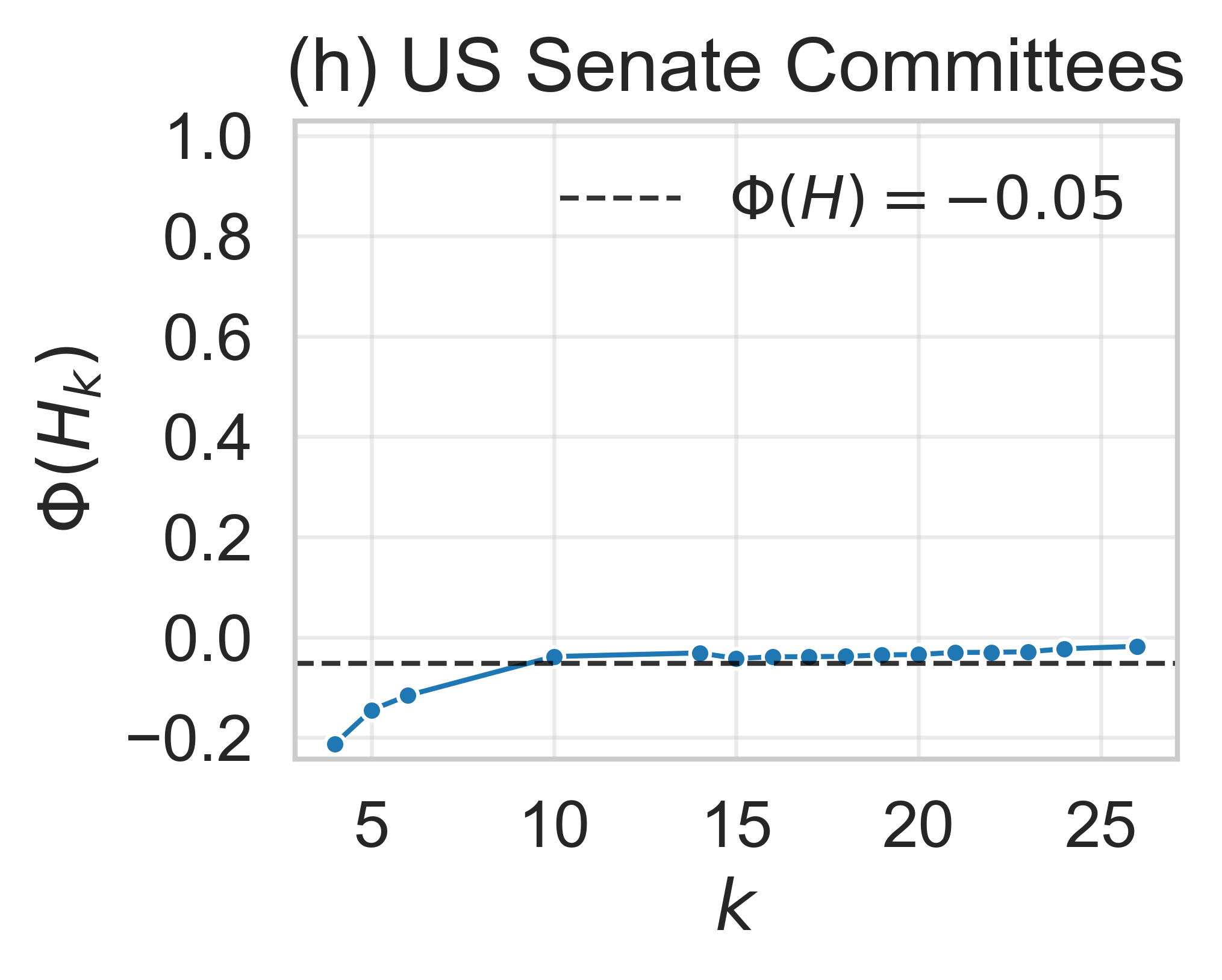}
    \end{subfigure}
    \vspace{-.5em}
    \caption{$\Phi(H_k)$ vs.\ $k$ across all datasets.}
    \label{fig:phi-vs-k}
\end{figure*}


We first analyze our method on synthetic hypergraphs generated using a Hypergraph Stochastic Block Model (HSBM) \cite{coleExactRecoveryHypergraph2020}. The HSBM produces a $k$-uniform hypergraph with $|M|$ equal-sized partitions, each of size $|V|/|M|$ (assuming $|V|$ is divisible by $|M|$). For any subset of $k$ nodes, it creates a hyperedge with probability $p$ if all nodes belong to the same partition, and with probability $q$ otherwise. 
For our experiments, we modify this model in such a way that it generates a pure hyperedge with probability $p$ when the homophilic parameter $p>0$, a balanced hyperedge with probability $|p|$ when $p<0$, and a random hyperedge otherwise. Fig.~\ref{fig:synthetic} presents results for a $10$-uniform hypergraph with $10$ attributes and $1000$ nodes. As shown in Fig.~\ref{fig:synthetic}(a), the Perplexity-Homophily Index $\Phi(H)$ increases linearly as $p$ varies from $-1$ to $+1$, matching the model’s intended structure. Fig.~\ref{fig:synthetic}(b) shows $\Phi(H)$ for different values of $k$ and $p$; the index accurately captures the intended homophily level.

Table~\ref{tab:datasets} summarizes the real-world datasets used in our analysis. For any hypergraph $H$, let $H_k$ be the sub-hypergraph of $H$ obtained by retaining only hyperedges of size $k$. In Fig. \ref{fig:phi-vs-k}, for each dataset $H$, we plot and compare $\Phi(H_k)$ for all $k$, where the black dashed line shows the value of $\Phi(H)$ for reference, and blue dots show the value of $\Phi(H_k)$ for each $k$. 
It shows that the homophily $\Phi(H_k)$ varies substantially with interaction size $k$. 
\begin{figure*}[t]
    \centering
    \begin{subfigure}[t]{0.24\textwidth}
        \centering
        \includegraphics[width=\linewidth]{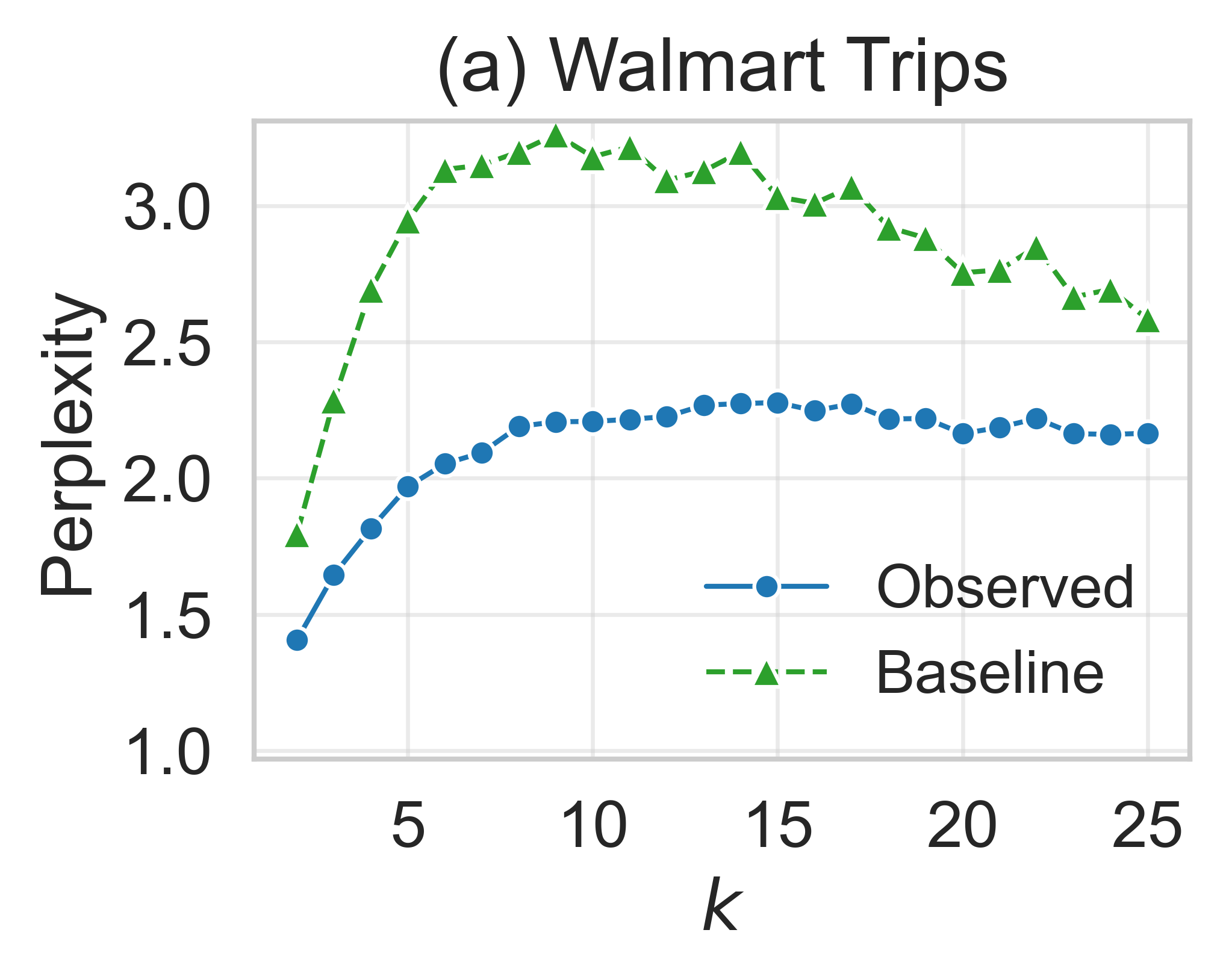}
    \end{subfigure}
    \begin{subfigure}[t]{0.24\textwidth}
        \centering
        \includegraphics[width=\linewidth]{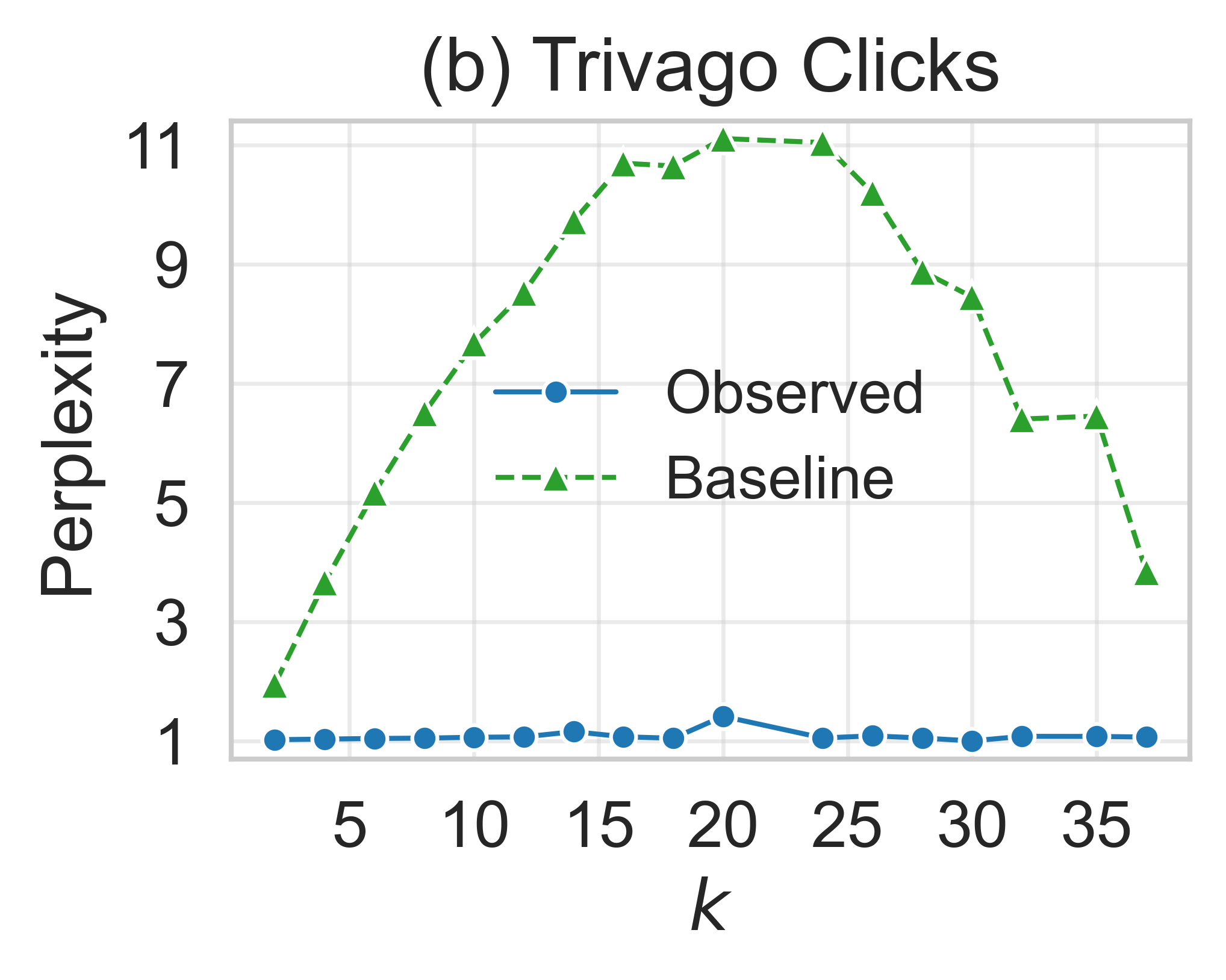}
    \end{subfigure}
    \begin{subfigure}[t]{0.24\textwidth}
        \centering
        \includegraphics[width=\linewidth]{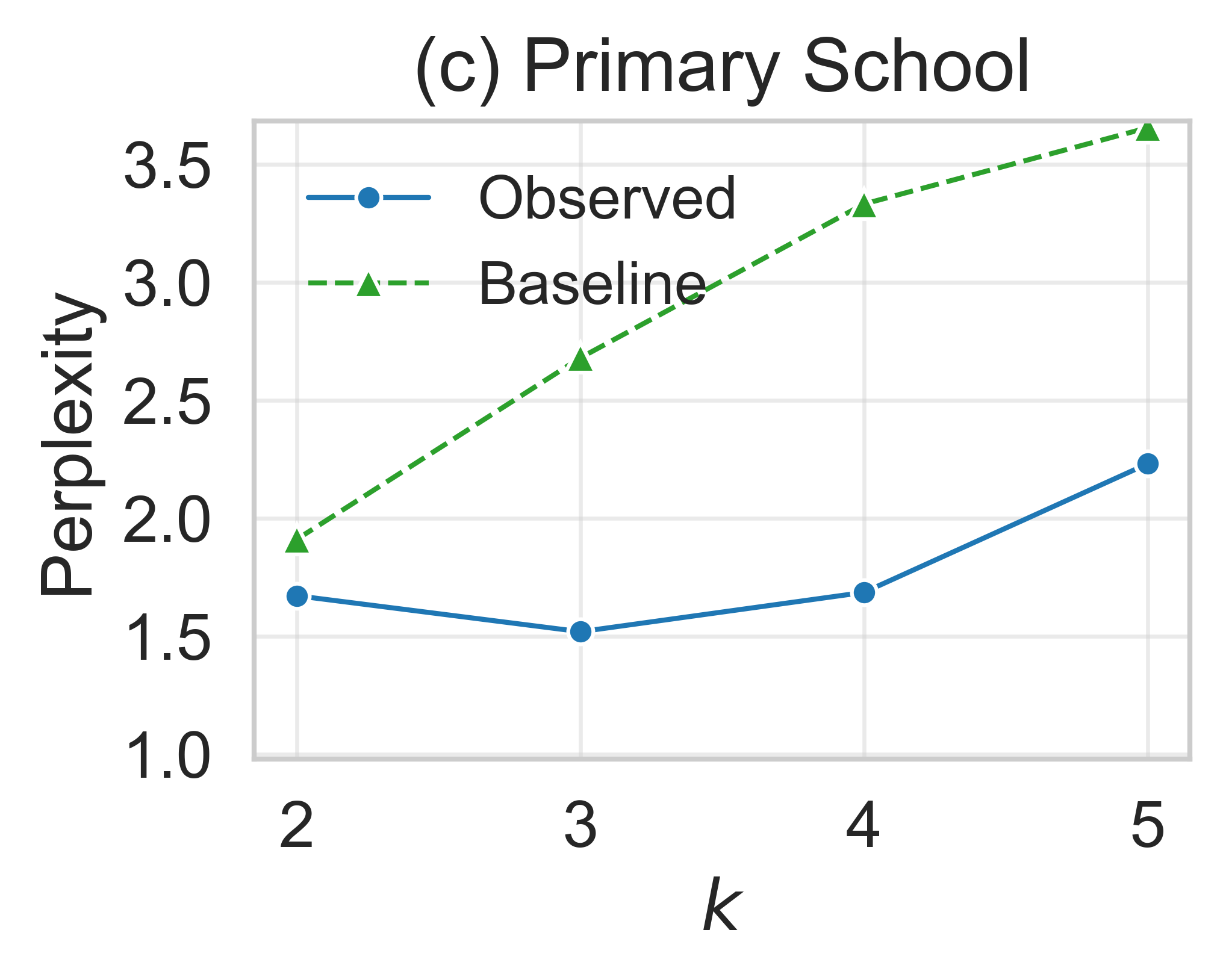}
    \end{subfigure}
    \begin{subfigure}[t]{0.24\textwidth}
        \centering
        \includegraphics[width=\linewidth]{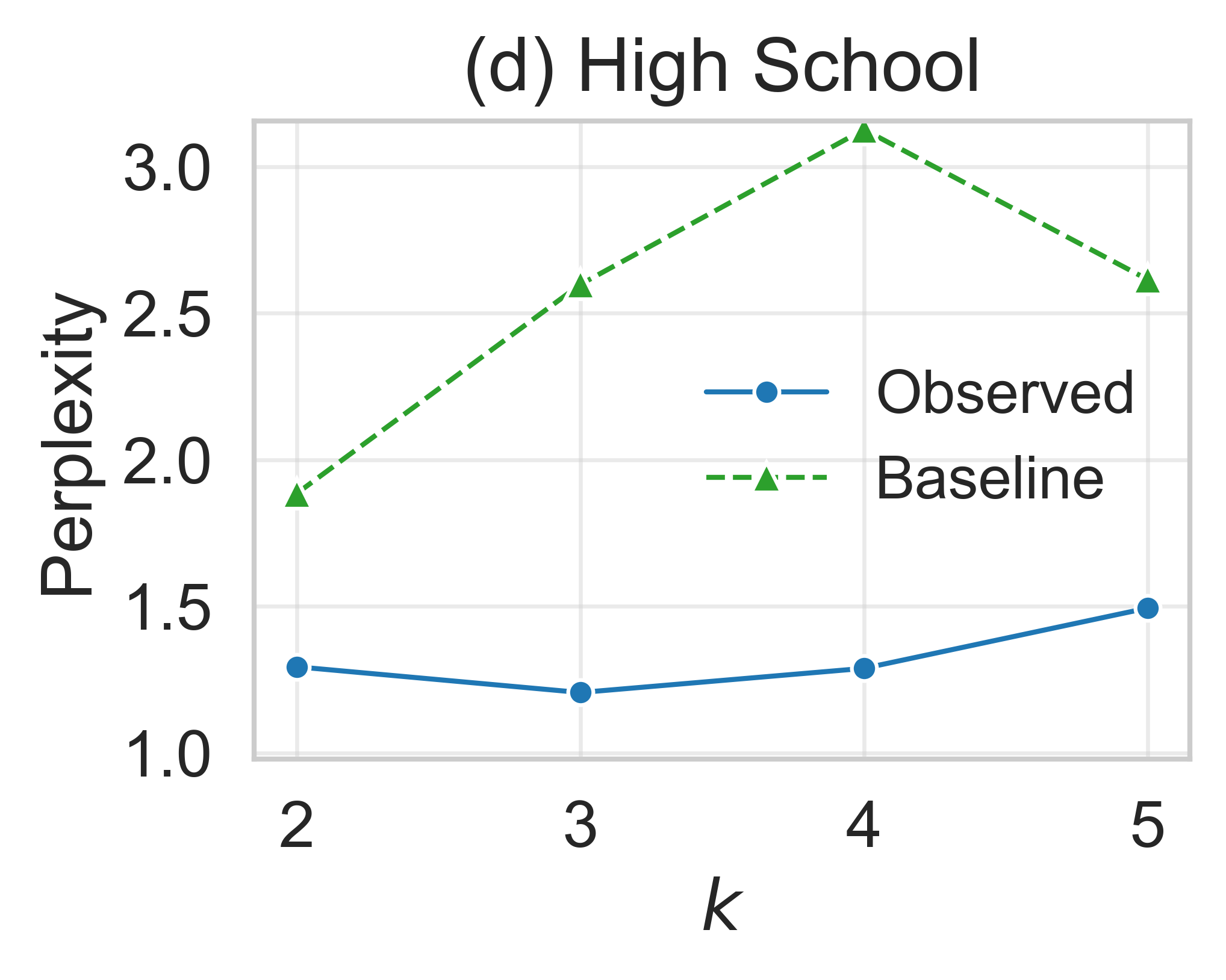}
    \end{subfigure}
    \begin{subfigure}[t]{0.24\textwidth}
        \centering
        \includegraphics[width=\linewidth]{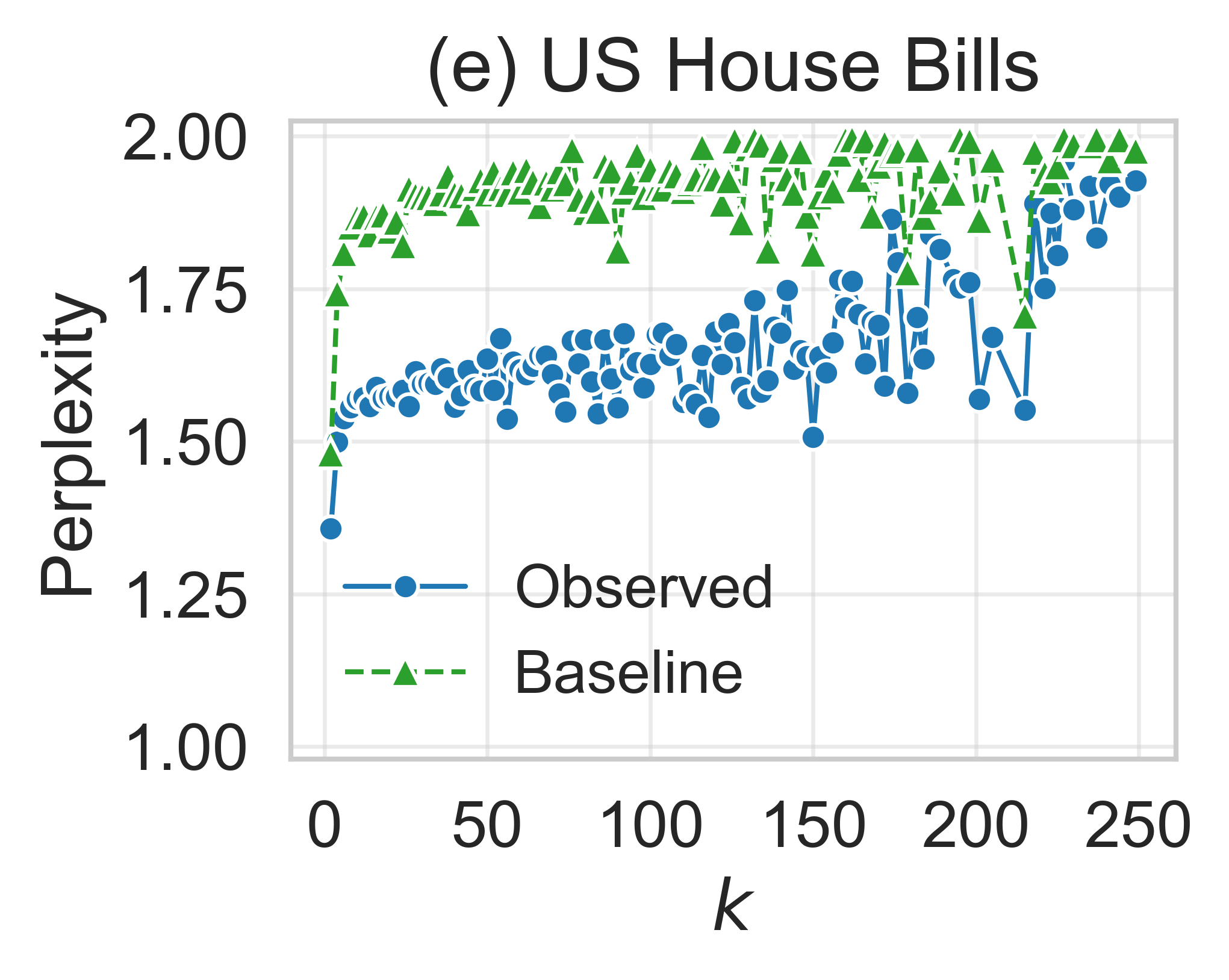}
    \end{subfigure}
    \begin{subfigure}[t]{0.24\textwidth}
        \centering
        \includegraphics[width=\linewidth]{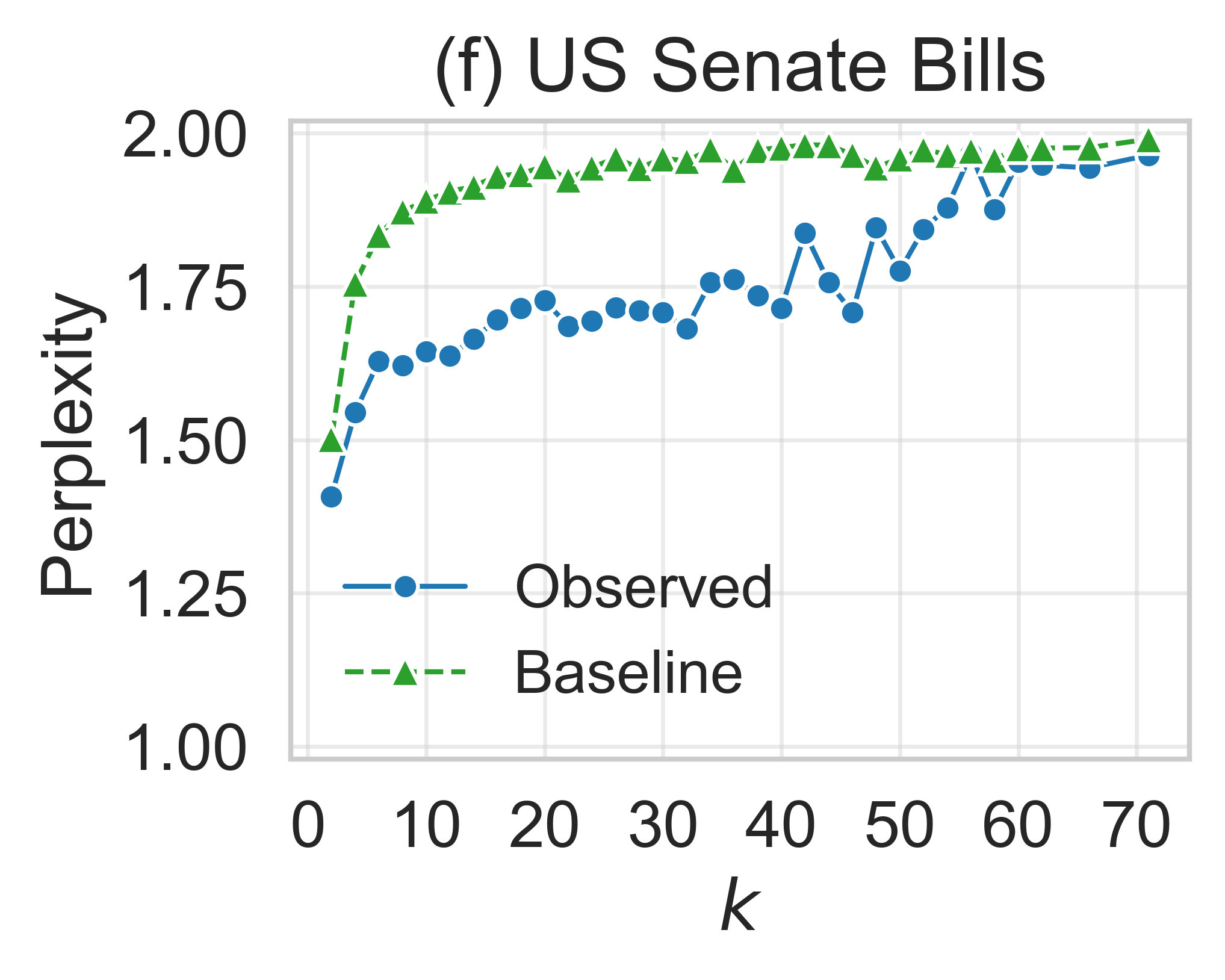}
    \end{subfigure}
    \begin{subfigure}[t]{0.24\textwidth}
        \centering
        \includegraphics[width=\linewidth]{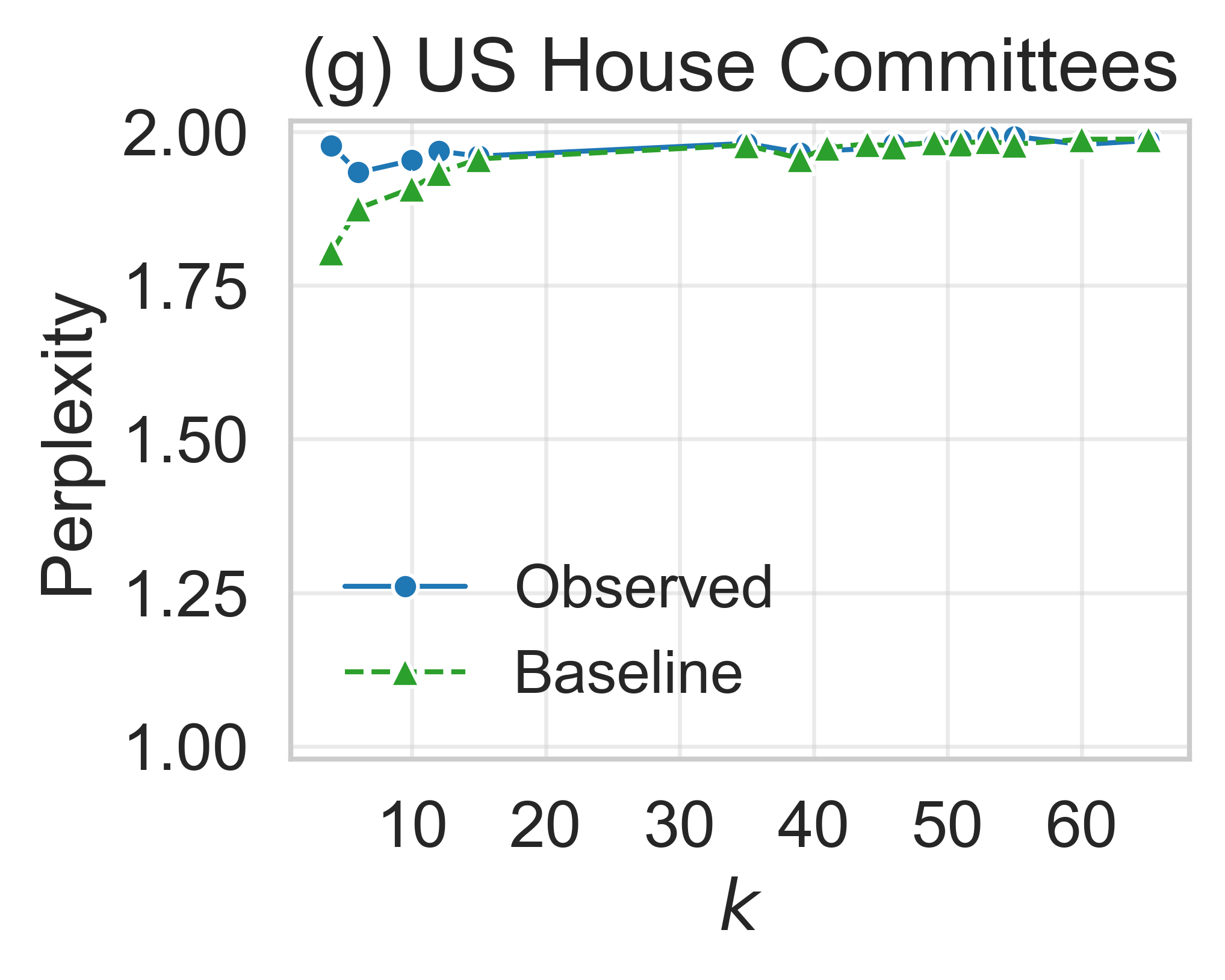}
    \end{subfigure}
    \begin{subfigure}[t]{0.24\textwidth}
        \centering
        \includegraphics[width=\linewidth]{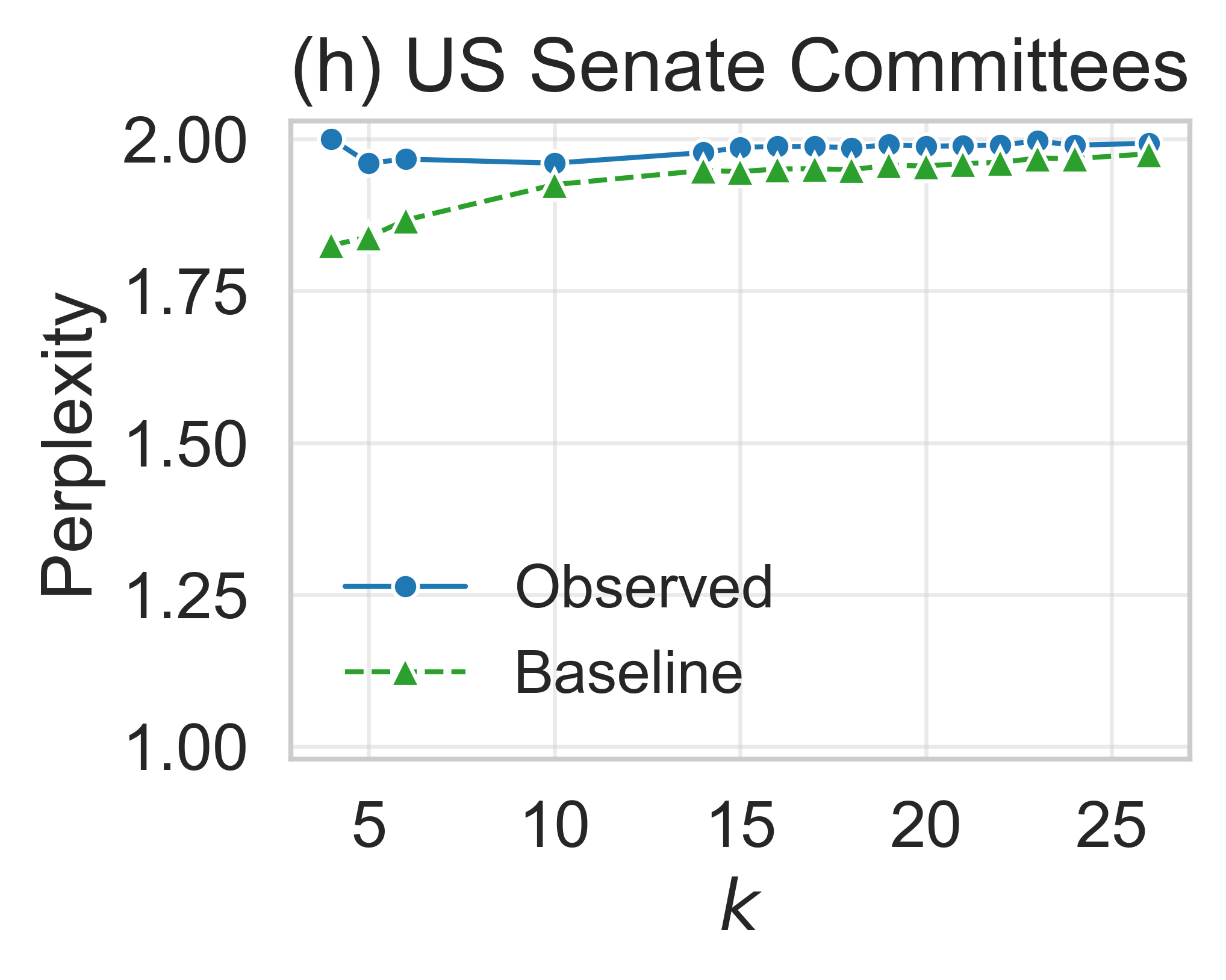}
    \end{subfigure}
    \caption{Perplexity vs. hyperedge size ($k$).}
    \label{fig:perp}
\end{figure*}
We first observe that the method accurately identifies extreme cases of homophily. The Trivago Clicks hypergraph exhibits near-perfect homophily with $\Phi(H)=0.98$, reflecting highly focused user browsing behavior. As shown in Fig.~\ref{fig:phi-vs-k}(b), $\Phi(H_k)\approx 1$ across all session lengths. Fig.~\ref{fig:perp}(b) compares the observed perplexity with the baseline perplexity for each value of $k$. Although the baseline perplexity varies with $k$, suggesting that a random session would check out hotels in multiple countries, the observed perplexity remains constant at $D(e)=1$. The decrease in baseline is due to the fact that a limited number of countries participate and are available for baseline calculation in large sessions. These findings confirm that users browse hotels exclusively within a single country, regardless of session length. In contrast, the US~Congress committees dataset exhibits homophily scores consistently close to zero (Figs.~\ref{fig:phi-vs-k}(g), \ref{fig:phi-vs-k}(h)), reflecting that the committees are intentionally constructed to maintain balanced representation across political parties. As a result, their attribute composition is identical (Fig.~\ref{fig:perp}(g)) or even more diverse from that of the null model (Fig.~\ref{fig:perp}(h)).

A strong dependence on interaction size emerges in datasets describing legislative co-sponsorship. In both the US~House and Senate, homophily decreases sharply once the number of co-sponsors exceeds a threshold (Figs.~\ref{fig:phi-vs-k}(e), \ref{fig:phi-vs-k}(f)). Small bills exhibit high homophily, whereas large bills have $\Phi \approx 0$, indicating that bipartisan support is necessary for attracting many co-sponsors. Correspondingly, the perplexity of large bills becomes indistinguishable from the null model (Figs.~\ref{fig:perp}(e), ~\ref{fig:perp}(f)). A similar but more gradual pattern appears in the Walmart shopping data: small carts are strongly homophilic, but homophily decreases with cart size (Fig.~\ref{fig:phi-vs-k}(a)). Observed perplexity saturates to a value of $\approx$\num{2.5} as $k$ increases, while baseline perplexity rises to \num{11} then falls back to \num{4}, making the largest carts match more closely with the null model (Fig.~\ref{fig:perp}(a)). %

Finally, in the high school and primary school contact networks, the strongest homophily occurs for groups of size $k=3$ and $k=4$, with lower homophily for pairs and larger groups. High school students show stronger class-based segregation than primary school students, reflected in the higher overall homophily ($\Phi(H)=0.73$ vs.\ $0.43$). The curve for baseline is consistently above the observed perplexity in Figs. \ref{fig:perp}(c) and \ref{fig:perp}(d). For all group sizes, the observed perplexity remains below $1.5$ for High School, while it stays above $1.5$ for Primary School.
Overall, these results show that the Perplexity-Homophily Index captures both overall homophily and its dependence on interaction size.

\section{Discussion and Conclusion}

The proposed framework provides a general and flexible approach for quantifying homophily in hypergraphs. By computing the diversity of each hyperedge, comparing it with the expected diversity under a degree- and attribute-aware null model, and defining homophily as a normalized deviation between the two, our method extends the concept of homophily well beyond the limitations of pairwise networks. Traditional graph-based measures, which only distinguish between same-attribute and different-attribute edges, are insufficient in higher-order settings where interactions may involve several nodes and multiple attributes. In contrast, the use of perplexity naturally incorporates both the number of distinct attributes and their proportions, making it well-suited for capturing the richness of higher-order interactions.

It is important to emphasize that perplexity is just one member of a larger family of diversity indices, specifically the order-$1$ Hill number \cite{hillDiversityEvennessUnifying1973}. The framework is not restricted to this choice. For example, the order-$2$ Hill number (the inverse Simpson index), widely known as the Laakso-Taagepera index to measure ``effective number of parties'' \cite{laakso1979effective} places greater emphasis on majority attributes and discounts minority ones. In an attribute composition such as $(8,1,1)$, the inverse Simpson index yields significantly lower diversity ($\approx1.52$) than perplexity ($\approx1.89$) because the squared term strongly amplifies the dominant class. By using perplexity, the present framework maintains sensitivity to the full distribution of attributes and provides a more balanced assessment of diversity within a hyperedge. 

Overall, the framework is adaptable and can be paired with any suitable diversity index, depending on the application and the desired sensitivity to minority attributes. This flexibility, together with its natural extension of homophily to higher-order structures, makes the framework broadly applicable for analyzing complex datasets where interactions extend beyond simple pairwise relations. 
Future research may extend this framework to temporal and multilayer hypergraphs and explore alternative diversity indices to tailor sensitivity to specific applications. Integrating the method with generative higher-order models would further enable parameter inference and realistic benchmark construction.




\bibliographystyle{ACM-Reference-Format}
\bibliography{references}

\end{document}